\documentclass[12pt,preprint2]{aastex}
\input epsf

\shorttitle{Magnetically Dominated Star Forming Regions}
\shortauthors{Lim, Falle, \& Hartquist }

\begin{document}

\title{The Production of Magnetically Dominated Star Forming Regions}


\author{A.J.   Lim\altaffilmark{1},   S.A.E.G.  Falle\altaffilmark{2},
T.W. Hartquist\altaffilmark{1}}


\altaffiltext{1}{School of Physics and Astronomy, University of Leeds,
Leeds LS2 9JT, UK.}
\altaffiltext{1}{Department  of  Applied  Mathematics,  University  of
Leeds, Leeds LS2 9JT, UK.}

\begin{abstract}
We consider the  dynamical evolution of an interstellar  cloud that is
initially  in  thermal equilibrium  in  the  warm  phase and  is  then
subjected to a sudden increase in the pressure of its surroundings. We
find  that if  the initial  plasma $\beta$  of the  cloud is  of order
unity, then there  is a considerable period during  which the material
in the cloud both has a small $\beta$ and is in the thermally unstable
temperature  range.  These  conditions  are not  only consistent  with
observations of star  forming regions, but are also  ideally suited to
the  production  of  density  inhomogeneities  by  magnetohydrodynamic
waves. The end result should be a cloud whose size and average density
are typical of Giant Molecular Clouds (GMCs) and which contains denser
regions whose densities are in  the range inferred for the translucent
clumps in GMCs.
\end{abstract}




\section{Introduction}

Observations of magnetic fields in star forming regions (e.g. Crutcher
1991) tell us that the plasma $\beta$  is of order $0.04$ and that the
velocity dispersion is  of the order of the  Alfv\'en speed.  For this
reason, many of the simulations  of the early stages of star formation
(e.g.   Fiedler \&  Mouschovias 1993;  Ballesteros-Paredes  \& Mac-Low
2002; Falle \& Hartquist 2002;  Padoan \& Nordlund 2002; Gammie et al.
2003) have assumed  that the initial  value of $\beta$ is  low.  These
papers  show   that  dense  structures  can  be   formed  under  these
conditions, but they do not consider  how $\beta$ came to be so low in
the first place.

On scales large compared to the separation between clouds, the thermal
and  magnetic pressures  in  the interstellar  medium are  comparable.
Locally  $\beta$  may be  about  $0.5$, while  in  the  midplane at  a
distance of  5 kpc  from the  Galactic Centre, it  may be  about $1.5$
(Ferriere  1998).   The  production  of  a star  forming  region  must
therefore  also reduce  $\beta$  by a  considerable  factor.  Here  we
examine  a  mechanism  for  this  which relies  upon  pressure  driven
compression and radiative cooling.

This  process produces  a  transient  structure, but  this  is not  in
conflict with  estimates of only a  few million years for  the ages of
the stars  in Giant  Molecular Clouds (GMCs)  (Ballesteros-Paredes, et
al. 1999;  Hartman 2003).   This is also  not very different  from the
time that  a clump moving at  the observed velocities of  $\sim 10$ km
s$^{-1}$ takes to  travel a significant fraction of the  size of a GMC
($\sim 100$ pc). We shall therefore assume that GMCs are transient and
that their formation is triggered by external disturbances. We make no
attempt to follow  the entire evolution of a  GMC; instead we describe
an idealised numerical calculation  that illustrates how a cloud might
make the transition from a state with moderate $\beta$ to one with low
$\beta$.

A significant  fraction of  interstellar space is  filled by  $10^6$ K
gas, which is composed of overlapping supernova remnants (Cox \& Smith
1974; McKee \& Ostriker 1977).  This gas is continually being reheated
by shocks  from supernova explosions. One estimate  (McKee \& Ostriker
1977) indicates  that, on  average, this  causes the  pressure  at any
given point to fluctuate significantly on timescales of the order of a
million  years or  less.  Such  pressure fluctuations  also  appear in
simulations   of   the    structure   of   the   interstellar   medium
(e.g. Gazol-Pati\~no \& Passot 1999; Korpi et al.  1999; de Avillez \&
Breitschwerdt  2004,  2005; Mac-Low  et  al.   2005).   de Avillez  \&
Breitschwerdt (2005)  and Mac-Low et  al. (2005) do, indeed,  find low
$\beta$  regions  in  their  simulations. However,  these  are  global
simulations  containing a  number of  ingredients, so  that it  is not
entirely  clear  how  these   regions  are  produced.   Our  idealised
calculation not  only has the  advantage that the mechanism  is clear,
but also makes  it possible to use higher  resolution than is possible
in  a global  simulation.  As  we shall  see later,  even  this higher
resolution  is   not  sufficient   to  follow  the   entire  evolution
accurately.

\section{The Model}

\subsection{Heating and Cooling}

For the heating and cooling  rates appropriate for diffuse atomic gas,
there  are  two thermally  stable  phases  which  can be  in  pressure
equilibrium with each other for a range of thermal pressures.  The low
density warm  phase is at $\sim  10^4$ K, while the  high density cold
phase  is at $\sim  10^2$ K.   If molecules  form, as  they do  in the
clumps  within GMCs,  then the  temperature of  the cold  stable phase
decreases.  The thermal pressure in the warm phase has a maximum above
which only the cold phase can be in thermal equilibrium.

As a  simple model  of the  thermal behaviour of  the material  in the
cloud we  adopt the following form  for the net heating  rate per unit
volume

\[
H = \rho [0.015  - \rho \Lambda(T)] \mbox{~~erg~cm$^{-3}$~s$^{-1}$}
\]

\noindent
where,

\[
\begin{array}{ll}
\Lambda(T) = 3.564~10^{16} T^{2.21} & 0 \le T < 141 \mbox{~K} \\
\Lambda(T) =  9.1~10^{18} T & 141 \le T < 313 \mbox{~K} \\
\Lambda(T) =  1.14~10^{20} T^{0.56} & 313\le T < 6102 \mbox{~K} \\
\Lambda(T) =  1.924 T^{3.67} & 6102 \le T < 10^5 \mbox{~K} \\
\Lambda(T) =   1.362 T^{-0.5} & T \ge 10^5 \mbox{~K} \\
\end{array}
\]

\noindent
(S\'anchez-Salcedo  et. al 2002).   $H$ is  of the  same form  as that
appropriate for the diffuse atomic  gas (Wolfire et al.  1995).  If we
ignore the dependence of  the ionization fraction on temperature, then
the pressure is

\[
p = {{\rho k T} \over m}
\]

\noindent
where $m =  2~10^{-24}$ gm is the mean particle  mass. For this model,
the  maximum  density  and  pressure  of the  warm  phase  in  thermal
equilibium are  $0.5$ cm$^{-3}$ and $3051k$ respectively.  Here $k$ is
Boltzmann's constant.

\subsection{Initial Conditions}

It is well known that a transition from the warm to the cold phase can
be triggered by a relatively  small pressure increase caused either by
flow convergence (Hennebelle \& P\'erault  1999, 2000) or a weak shock
(Koyama  \& Inutsuka  2000, 2002,  2004). If  $\beta$ is  initially of
order  unity,  then this  transition  can  produce  a small  value  of
$\beta$. We therefore consider a  cloud which is initially in the warm
phase  and  in  pressure   equilibrium  with  the  surrounding  hotter
material.   The hot  gas  can be  assumed  to be  adiabatic, but  both
heating and radiative cooling are  important within the cloud. We then
suppose that the cloud is subjected  to an increase in the pressure of
the  surrounding hot material  from a  value at  which the  warm phase
exists to one at which it  does not.  The pressure increase is assumed
to  take place  on a  timescale  that is  short compared  to the  time
required for the cloud to adjust.

In all the calculations, the cloud is initially in thermal equilibrium
in  the  warm  phase  with  a  number  density  of  $0.45$  cm$^{-3}$,
corresponding to a temperature of  $6277$ K and a pressure of $2825k$.
The external pressure is imposed by embedding the cloud in uniform hot
material with density $0.01$  cm$^{-3}$ with the appropriate pressure.
This material is  not in thermal equilibrium, but  its cooling time is
so long that  it can be assumed to  behave adiabatically.  The initial
magnetic field  is in  the axial direction  and is  uniform everywhere
with  a pressure  equal  to the  thermal  pressure in  the cloud  i.e.
$\beta = 1$ in the cloud.  Since the cloud is initially spherical, the
flow  remains axisymmetric. Table  1 lists  the external  pressure and
size of the  cloud for the five cases  considered.  Changing the cloud
radius only  affects the ratio of  cloud radius to  the cooling length
behind the fast mode shock  that propagates into the cloud. This ratio
is $\simeq 0.1$ for a cloud radius of $200$ pc. Note that the external
thermal  pressures   are  well  within  the  range   found  in  global
simulations.

\begin{table}
\begin{center}
\begin{tabular}{lcc}
 & External Gas Pressure & Cloud Radius \\
Model A & $15000 k$ & $200$ pc \\ 
Model B & $10000 k$ & $200$ pc \\
Model C & $10000 k $ & $50$ pc \\
Model D & $7500 k$ & $200$ pc \\
\multicolumn{3}{c}{(decreasing external pressure)} \\
Model E & $10000 k$ & $200$ pc
\end{tabular}
\caption{External pressure and cloud radius.}
\end{center}
\end{table}

Model E  explores the effect  of an external thermal  pressure, $p_e$,
that decreases with time according to

\[
p_e = 
\left\{
{
\begin{array}{ll}p_0 e^{-t/t_c} & \mbox{for~$p_e > p_0$} \\
p_0 & \mbox{otherwise.} \\
\end{array}
}
\right.
\]

\noindent
Here $p_0 = 10000 k$ is the initial external thermal pressure and $t_c
= 2.6~10^7$  yrs, which is approximately  twice the time  it takes the
fast  shock  to  return  to   the  edge  of  the  cloud  (see  section
\ref{results}) .

\subsection{Computational Details}

All the  calculations were done in axisymmetry  using the unstructured
hierarchical adaptive  grid code described in Falle  \& Giddings 1993.
This uses  a hierarchy of  grids $G^0 \cdots  G^N$ such that  the mesh
spacing on grid $G^n$ is $\Delta x_0/2^n$. Grids $G^0$ and $G^1$ cover
the  whole domain,  but  the finer  grids  only exist  where they  are
needed. The solution at each  position is calculated on all grids that
exist  there and  the difference  between these  solutions is  used to
control refinement.  In order to ensure Courant number matching at the
boundaries between coarse  and fine grids, the timestep  on grid $G^n$
is  $\Delta t_0/ 2^n$  where $\Delta  t_0$ is  the timestep  on $G^0$.
Such a  grid structure is very  efficient for flows  that contain very
thin dense regions since it confines  the fine grids to where they are
needed.  The basic algorithm  is an  MHD version  of the  second order
Godunov scheme described in Falle  (1991).  This uses a linear Riemann
solver and  the divergence cleaning  algorithm described in  Dedner et
al. (2002).
 
The computational domain  is $0 \le r  \le 2 R_c$, $0 \le  z \le 2R_c$
where $R_c$ is the cloud  radius. The boundary conditions are symmetry
on the $z = 0$ plane  and the axis and free flow conditions elsewhere.
The low density in the external  gas ensures that the boundaries at $r
= 2 R_c$ and $z = 2 R_c$ have a neglible effect on the solution. There
were $6$  grid levels with the  finest being $1280  \times 1280$. This
gives a mesh spacing of $0.31$ pc  for models A, C, D, E and $0.08$ pc
for model B.  Even though this is significantly higher resolution than
that attained by de Avillez \& Breitschwerdt (2005) and Mac-Low et al.
(2005) it is  not sufficient to  follow the complete evolution  of the
cloud.  It is, however, enough  to show the formation of the fragments
that we identify with the  translucent clumps and to give some insight
into their initial evolution.

\subsection{Results}
\label{results}

Figure 1 shows  grey-scale maps of the number  density, magnetic field
lines and $\beta$  for model A at the times shown.   A fast mode shock
is driven  into the  cloud, which reflects  off the symmetry  axis and
returns to the  edge of the cloud  at the last time shown  in Figure 1
($13.2~10^6$ yrs).   At this point most  of the cloud  has $\beta \sim
0.03$ and is  in the thermally unstable temperature  range.  Note that
self-gravity can only have a minor effect upon the dynamics up to this
point since even at the latest time shown in the Figure, the free-fall
time is $\simeq 3~10^7$ yrs,  whereas the crossing time at the thermal
sound speed is $\simeq 2~10^7$ yrs  and that at the fast mode speed is
$\simeq 6~10^6$ yrs.

At this point the mean density  of the cloud is $\simeq 30$ cm$^{-3}$,
which  is typical  of that  in a  GMC. Furthermore,  the low  value of
$\beta$ provides  the ideal conditions  for the generation  of density
inhomogeneities  by  magnetohydrodynamic  waves  (Falle  \&  Hartquist
2002), particularly since most of the gas is in the thermally unstable
temperature  range. The  end result  must  be the  formation of  dense
regions that  are in the  cold phase embedded  in diffuse warm
phase.  Indeed,  Figure 1 shows  that such dense regions  have already
begun  to form  at the  latest  time shown.  Unfortunately, we  cannot
follow the evolution  of these dense regions much  beyond this because
not  only  does  the   numerical  resolution  become  inadequate,  but
self-gravity  begins to  influence  the large  scale  dynamics of  the
cloud.  However, it  is clear that these regions  will eventually come
into approximate pressure  equilibrium with the rest of  the cloud, in
which case their densities must  be in the range $500-1000$ cm$^{-3}$.
This is  typical of  the densities in  the translucent clumps  in GMCs
(e.g.  Williams, et al. 1995).

There is also a slow mode  shock which moves much slowly than the fast
mode shock  because the reduction  in the magnetic pressure  across it
means there is  nothing to stop the gas being  compressed as it cools,
so that  it comes  into thermal equilibrium  in the cold  phase.  This
shock therefore produces  a thin dense shell near  the boundary of the
cloud in which  $\beta \simeq 1$.  Since the post-shock  gas is in the
thermally  unstable regime,  this  process is  unstable  and we  would
expect the  thin shell to  fragment.  Our calculations show  that this
does indeed occur at later times.

As one would expect, the  increase in magnetic field inhibits collapse
perpendicular to the  field so that the cloud  becomes flattened along
the  field  lines.   The  slow  shock must  eventually  eliminate  the
difference between  the thermal pressures in the  surroundings and the
cloud so  that it ends up  as a thin disc  in which $\beta$  is of the
same order as  that in the surroundings. However,  this process is not
complete until $t \simeq 2.4~10^7$ yrs.

The behaviour of the other  simulations is similar, except that larger
external  pressures lead  to faster  evolution and  smaller  values of
$\beta$. In order  to quantify this, in Figure 2  we show the fraction
of the volume of the cloud for  which $\beta < 0.1$.  It is clear that
in all cases there is an extended period during which $\beta < 0.1$ in
much of the volume of the cloud.  Figure 3 shows that for models A and
B, there is also a significant fraction of each cloud for which $\beta
<  0.05$.   The  results for  model  E  show  that  there is  still  a
reasonable  volume of  low  $\beta$ material  even  when the  external
thermal pressure decreases with  time, although the timescale for this
decrease must not  be much smaller than that in  model E. For example,
there is  little low  $\beta$ material when  $t_c$ is reduced  to half
this value.

Although  a  significant fraction  of  the  volume  of each  cloud  is
occupied by material with low $\beta$, the corresponding mass fraction
is somewhat lower.  This is because, once the  fast shock has returned
to the cloud boundary, the cloud consists of three regions: the region
interior to the slow shock, in which  $\beta$ is low and the gas is in
the thermally unstable phase; the  thin shell behind the slow shock in
which  $\beta \simeq  1$ and  the gas  is in  the cold  phase;  a thin
annulus on the  symmetry plane in which the  conditions are similar to
those  in the thin  shell.  Although  the thin  shell and  the annulus
contain a significant  fraction of the mass, they  only occupy a small
fraction of the volume.

The annulus  is not very  significant since it  is an artifact  of the
symmetry  of  the calculation  and  would not  be  present  in a  more
realistic  calculation without this  symmetry.  On  the other  hand, a
thin shell  must appear in  any such calculation. This  is interesting
since  it  is likely  to  break up  into  dense  fragments (Koyama  \&
Inutsuka 2000) and is therefore a possible site of star formation.  In
a more  realistic simulation  which is asymmetric  due to  an external
flow, it would not be as regular as in our simulation and so one might
expect star formation  to occur on one side of the  cloud.  There is a
good deal of observational evidence that suggests that high-mass stars
tend to form at the surfaces  of GMCs ((e.  g.  Israel 1987; Gatley et
al.   1979: Fich  et  al. 1982).   In  the case  of  the Galaxy,  this
evidence  is not conclusive,  but observations  of other  galaxies may
provide  clearer  results.   For  example,  there  is  some  tentative
evidence from M33 that young high-mass stars are slightly off-set from
the molecular material (Pattison \& Hoare 2005).

\section{Conclusions}

The  above  simulations  show  the formation  of  highly  magnetically
dominated  regions  in thermally  unstable  material behind  fast-mode
shocks  propagating through  a  cloud  that is  initially  in a  warm,
thermally stable  state.  The external pressure  variation required to
generate  such  regions is  well  within  the  range expected  in  the
mid-plane  of the  Galaxy.  A  slow-mode shock  follows  the fast-mode
shock leading to the production  of dense regions in which the thermal
and magnetic pressures are comparable.  These dense regions may be the
sites of  high-mass star  formation.

We thank L.  Blitz and M. Hoare for input and  the referee for helpful
comments  on the original  version.  A.  J.  Lim  acknowledges support
from PPARC during the course of this work.

\section{Figures}

Figure 1: Linear  grey scale plots of the  number density and magnetic
field lines  (left) and $\beta$ (right)  at the times  shown.  For the
number density the  range is $0$ --  $4$ and for $\beta$ it  is $0$ --
$1$.

Figure 2: Fraction of the volume of the cloud for which $\beta < 0.1$.

\medskip

Figure 3: Fraction of the volume of the cloud for which $\beta < 0.05$.

\end{document}